\documentclass{conm-p-l}

\copyrightinfo{2002}%            % copyright year
{American Mathematical Society}% copyright holder

\usepackage{fullpage,mathptm,amsmath,mathdots,amssymb,epsf,graphicx}
\advance\parskip 4pt
\usepackage{epstopdf}

\newtheorem{theorem}{Theorem}[section]
\newtheorem{lemma}[theorem]{Lemma}
\newtheorem{proposition}[theorem]{Proposition}
\newtheorem{corollary}[theorem]{Corollary}

\theoremstyle{definition}
\newtheorem{definition}[theorem]{Definition}

\theoremstyle{remark}
\newtheorem{remark}[theorem]{Remark}

\def\proof{\par\kern-\medskipamount\noindent\textbf{Proof.}~~}

\def\remark{\par\kern\medskipamount\noindent\textbf{Remark~\stepcounter{theorem}\arabic{section}.\arabic{theorem}}~}

\catcode`\@ 11
\@addtoreset{equation}{section}
\long\def\@makecaption#1#2{\vskip\abovecaptionskip
  \sbox\@tempboxa{\small #1: #2}%
  \ifdim \wd\@tempboxa >\hsize \small #1: #2\par
  \else \global \@minipagefalse \hb@xt@\hsize{\hfil\box\@tempboxa\hfil}\fi
  \vskip\belowcaptionskip}
\newenvironment{example}{\refstepcounter{theorem}
  \par\kern\medskipamount\noindent\textbf{Example~\thetheorem}~}{\par\smallskip}
\newenvironment{property}{\refstepcounter{theorem}
  \par\kern\medskipamount\noindent\textbf{Property~\thetheorem}~}{\par\smallskip}
\newenvironment{condition}{\refstepcounter{theorem}
  \par\kern\medskipamount\noindent\textbf{Condition~\thetheorem}~}{\par\smallskip}

\numberwithin{figure}{section}
\catcode`\@ 12

\let\==\bar
\let\@=\mathbf
\newcommand{\partialderiv}[3][]{\frac{\partial^{#1}#2}{\partial {#3}^{#1}}}
\newcommand{\ds}{\displaystyle}
\def\circ{\ifmmode\mathchar"220E\else$\mathchar"220E$\fi}
\def\Wr{\mathop{\mathrm{Wr}}\nolimits}
\def\diag{\mathop{\rm diag}\nolimits}
\def\Real{{\mathbb{R}}}
\def\sech{\mathop{\rm sech}\nolimits}
\def\half{{\textstyle\frac12}}
\def\rank{\mathop{\rm rank}\nolimits}
\let\next=\phi\global\let\phi=\varphi\global\let\varphi=\next
 
\renewcommand\labelitemi{\ifmmode\circ\else$\circ$\fi}
\allowdisplaybreaks

\begin{document}
\title{\bf A generating function for the $N$-soliton solutions of
the Kadomtsev-Petviashvili II equation}
\author{Sarbarish Chakravarty}
\address{Department of Mathematics, University of Colorado, Colorado Springs, CO 80933}
\email{chuck@math.uccs.edu}
\thanks{Partially supported by National Science Foundation Grant No. DMS-0307181}

\author{Yuji Kodama}
\address{ Department of Mathematics, Ohio State University, Columbus, OH 43210} 
\email{kodama@math.ohio-state.edu}
\thanks{Partially supported by National Science Foundation Grant No. DMS-0404931} 

\subjclass{37K10i,33D45i,05A15}

%\date{}

\begin{abstract}
This work describes a classification of the $N$-soliton solutions of
the Kadomtsev-Petviashvili II equation in terms of chord diagrams
of $N$ chords joining pairs of $2N$ points. The different classes of $N$-solitons
are enumerated by the distribution of crossings of the chords.
The generating function of the chord diagrams is expressed as
a continued fraction, special cases of which are moment generating
functions for certain kinds of $q$-orthogonal polynomials. 
\end{abstract}

\maketitle

%\tableofcontents
%\thispagestyle{empty}
%\newpage
%%%%%%%%%%%%%%%%%%%%%%%%%%%%%%%%%%%%%%%%%%%%%%%%%%%%%%%%%%%%%%%%%%%%%%%%%%%%%%%
\section{Introduction}
\label{s:introduction}
The classical theory of enumerative combinatorics has indeed a far-reaching
scope, encompassing disparate areas in mathematical, physical, biological and
social sciences. Combinatorial entities such as permutations, partitions, trees,
lattice paths, graphs and their various enumerations find applications ranging
from econometrics, DNA structures, and statistical mechanics to coding theory, knots
and enumerative algebraic geometry. The purpose of the present note is to elaborate
on a somewhat unexpected relationship between a classical combinatorial problem 
studied by Touchard in the 1950s and the classification of a special class of
solitary wave solutions ({\em solitons}) of an exactly solvable nonlinear partial
differential equation discovered some 20 years later. This nonlinear wave equation,
known after its discoverers as the Kadomtsev-Petviashvili (KP) equation,
\begin{equation}
\partialderiv{}x\left(-4\partialderiv ut
  +\partialderiv[3]ux +6u\partialderiv ux \right)
            + 3 \sigma\partialderiv[2]uy =0\,,
\label{e:KP}
\end{equation}
describes the evolution of small-amplitude, weakly two-dimensional
solitary waves in a weakly dispersive medium \cite{SovPhysDoklady15p539}.
Depending on the sign of $\sigma$, there are two versions of the KP 
equation namely, KPI and KPII. Throughout this article we consider
Eq.\eqref{e:KP} with $\sigma = + 1$, which is the KPII equation. The
function $u=u(x,y,t)$ is the rescaled amplitude of the
wave-form. The KP equation arises in many physical
settings including water waves and plasmas (see e.g. ~\cite{InfeldRowlands} for 
a review). It is a completely integrable system with remarkably rich
mathematical structure which is well-documented in several monographs
~\cite{AblowitzClarkson,Hirota,NMPZ1984}.

\begin{figure}[h!]
\centering
\raisebox{0.85in}{(a)}\includegraphics[scale=0.53]{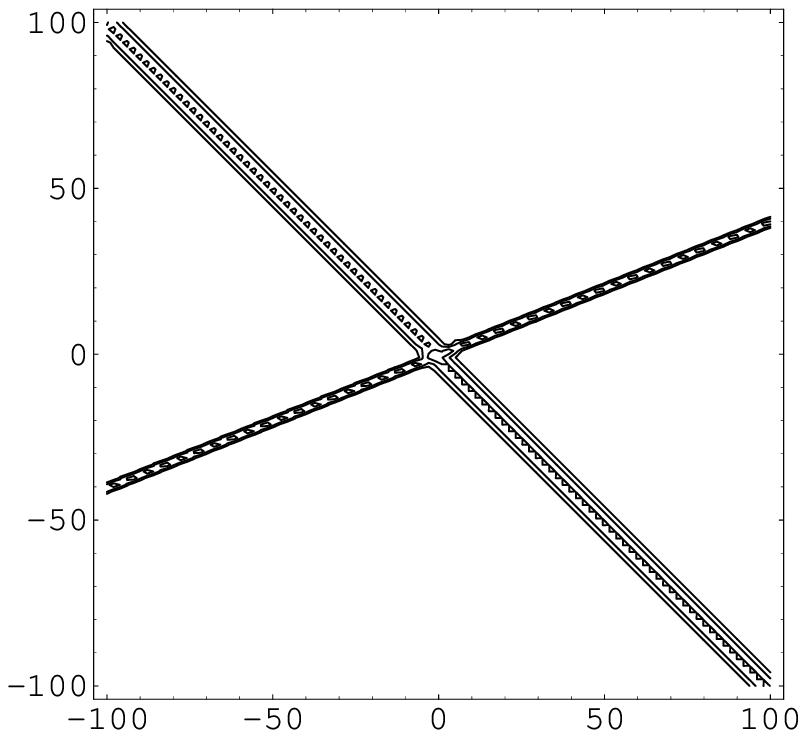} \hskip 0.3cm
\raisebox{0.85in}{(b)}\raisebox{-0.1cm}{\includegraphics[scale=0.53]{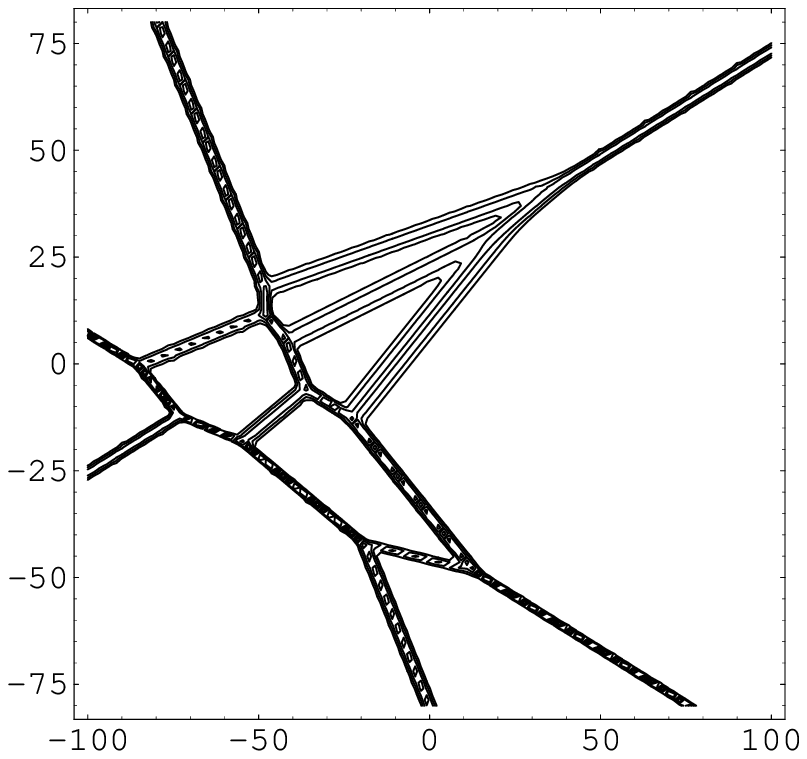}}\hskip 0.5cm
\raisebox{0.85in}{(c)}\includegraphics[scale=0.53]{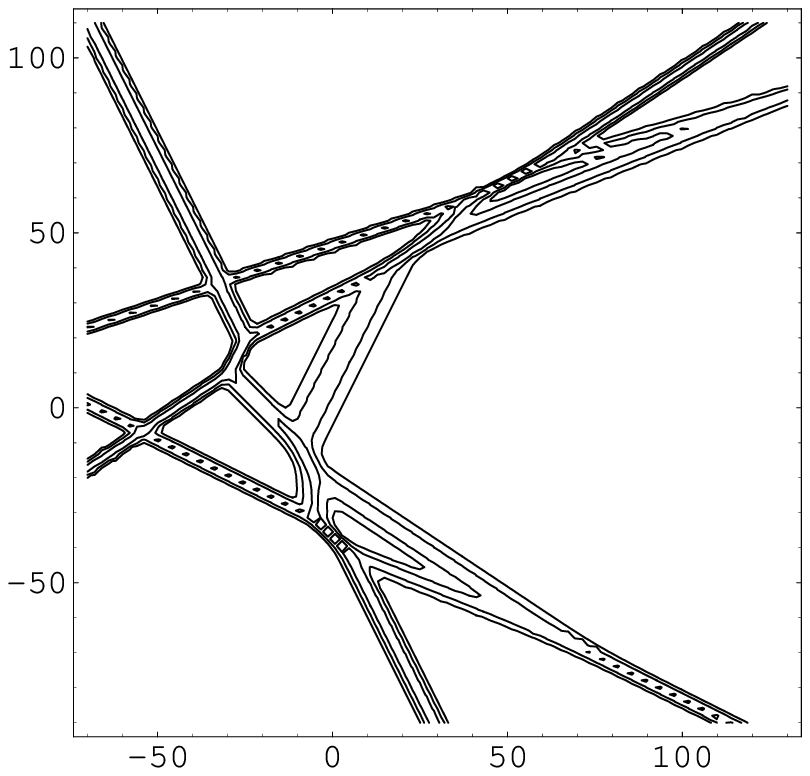}
\caption{N-soliton solutions of the KPII equation illustrating
different spatial interaction patterns:
(a) 2-soliton solution,
(b) resonant 3-soliton solution,
(c) partially resonant 4-soliton solution.
Here and in all following figures, the horizontal axis is $x$, vertical axis is $y$,
and the graphs show contour lines of $\ln u(x,y,t)$ for fixed~$t$.}
\label{f:kpfig}
\end{figure}

In this article, we consider a family of real, non-singular solitary wave
solutions of the KPII equation, known as the $N$-soliton solutions.
At any given time $t$, these wave-forms are localized along certain lines in 
the $xy$-plane, and decay exponentially everywhere else. In the generic case, 
they form a pattern of $N$ intersecting straight lines as $|y| \to \infty$
in the $xy$-plane, whereas in the near-field region the $N$ lines interact
to form intermediate lines and web-like structures as shown in Fig.~\ref{f:kpfig}.
The simplest kind of such solution is the 1-soliton, which is a constant amplitude
wave localized along a line in the $xy$-plane, and traveling with uniform velocity
perpendicular to the line. For $N > 1$, the asymptotic form of the $N$-soliton
solution coincides with $N$ 1-solitons along different directions, as $|y| \to \infty$ 
and uniformly in $t$. For this reason, these solutions are often referred to
as the {\em line-solitons}.

Several researchers~\cite{Freeman,PLA95p1,JPSJ1983v52p749,pashaev,Soomere} as well as
the authors~\cite{jphysa36p10519,Kodama,BC,BCa,mpag} have studied the soliton 
solutions of KPII. The general line-soliton configurations called
the $(N_-,N_+)$-solitons consist of  $N_-$ line solitons as $y \to -\infty$ and 
$N_+$ line solitons as $y \to \infty$~\cite{BC,mpag}. The $N$-solitons 
correspond to the special case when $N_- = N_+ = N$, and when the $N$ line solitons 
as $y \to \infty$ are pair-wise identical (as wave-forms) to the $N$ line solitons 
as $y \to -\infty$. An interesting feature of the $N$-soliton solutions is the 
fact that these solutions can be essentially (i.e., up to space-time translations)
reconstructed from the asymptotic data alone, comprising $N$ pairs of
amplitudes and directions associated with the $N$ line solitons 
as $|y| \to \infty$~\cite{BCa,mpag,CK}.  As a direct consequence of this result, 
it is possible to classify all the $N$-soliton solutions into $(2N-1)!!$ distinct equivalence 
classes, corresponding to the ways of partitioning the integer 
set $\{1,2,\ldots,2N\}$ into $N$ distinct pairs. The purpose of this paper is to
extend our studies to a characterization of the $N$-soliton solutions according to 
soliton interaction patterns, and give a classification of such interactions in terms 
of certain partitions (perfect matchings) of the integer set $\{1,2,\ldots,2N\}$.
We highlight some interesting connections between $N$-solitons of KPII on one hand,
and combinatorics of chord diagrams and $q$-Hermite polynomials on the other.
\kern-\medskipamount

%%%%%%%%%%%%%%%%%%%%  Background  %%%%%%%%%%%%%%%%

\section{Background}
In this section we give a brief overview of the chord diagrams with $2N$ points,
as well as the line-soliton solutions of the KPII equation. The aim is to underscore
the connection between these two seemingly disjoint mathematical objects. In particular,
we illustrate how a KPII line-soliton can be represented by a set of index pairs, which
leads naturally to the construction of a chord diagram on an integer set.
We shall use this construction in Section~\ref{s:elastic}, to identify each line soliton 
of an $N$-soliton solution by using a chord joining two specific points among $2N$ points,
and study the soliton interaction patterns in terms of such chord diagrams.
\label{s:background}
\subsection{Chord diagrams}
\begin{figure}[h!]
\centering
\includegraphics[scale=0.48]{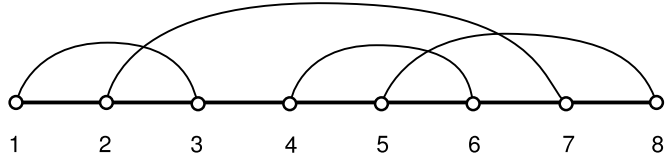}
\caption{A linear Chord diagram}
\label{f:chord1a}
\end{figure}

Let us first describe a chord diagram consisting of $N$ chords.
Consider a partition of the integer set $[2N] := \{1,2,\ldots,2N\}$ into
$N$ distinct $2$-element blocks or {\em pairings}
$$ {\bf p}_n := [i_n, \, j_n]\,, \quad 1 \leq i_n < j_n \leq 2N\,, \quad n=1,2,\ldots,N \,,$$
such that $[2N]$ is a union of the blocks ${\bf p}_1, {\bf p}_2, \ldots, {\bf p}_N$.
In combinatorics, such a partition is referred to as a (perfect) {\em matching} of $[2N]$. 
We will denote the set of all matchings of $[2N]$ by $\mathcal{M}_N$. The total number
of matchings in $\mathcal{M}_N$ is given by
$$|\mathcal{M}_N| = 1 \cdot 3 \cdot 5 \ldots \cdot (2N-1) =: (2N-1)!! \,.$$
A standard way to represent a matching $X$ of $\mathcal{M}_N$, is to mark $2N$ points
on a line from left to right labeled by $1,2,\ldots,2N$, and join the two points
 of each pairing $\bf{p}$ by a semicircular arc above the line. The resulting diagram
(see e.g. Fig.~\ref{f:chord1a}) is called a {\em linear} chord diagram,
whereas a chord diagram 
would correspond to labeling the $2N$ points in a clockwise manner on a circle,
and joining the two points of each pairing by a chord.

Without loss of generality, the smallest integer $i_n$ from each pairing
of $X \in \mathcal{M}_N$ can be arranged in a strictly increasing order 
$1=i_1 < i_2 < \ldots i_N \leq 2N-1$. However the $j_n$'s are not ordered in general.

\begin{definition} 
Let ${\bf p}_r$ and ${\bf p}_s$ with $r < s$ be distinct pairings 
(equivalently, a pair of chords). Then,
\begin{itemize} 
\item[(a)]${\bf p}_r$ and ${\bf p}_s$ form an {\em alignment} or a O-type configuration
if $i_r < j_r < i_s < j_s$. That is, the pairs do not overlap.
\item[(b)] 
${\bf p}_r$ and ${\bf p}_s$ form a {\em crossing} or a T-type configuration
if $i_r < i_s < j_r < j_s$. That is, the pairs partially overlap.
\item[(c)] ${\bf p}_r$ and ${\bf p}_s$ form a {\em nesting} or a P-type configuration
if $i_r < i_s < j_s < j_r$. That is the pairs completely overlap.
\end{itemize}
\label{d:config}
\end{definition}
The meanings of O-, T-, and P-type of configurations will be clear later in
Section~\ref{s:elastic} when we discuss the $N$-solitons of KPII.
In Fig.~\ref{f:chord1a}, the pairing $[4,6]$ forms an alignment 
(O-type configuration) with $[1,3]$; a crossing with $[5,8]$; and a nesting  
with the pairing $[2,7]$. Furthermore, the total number of (pairwise) crossings
in Fig.~\ref{f:chord1a} is 3; they occur between the pairs $[1,3]$ and $[2,7]$; 
$[2,7]$ and $[5,8]$; and $[4,6]$ and $[5,8]$. Similarly, there is one nesting,
$\{[2,7],[4,6]\}$, and two alignments, $\{[1,3],[4,6]\}$ and $\{[1,3],[5,8]\}$.

It should be clear that the of alignments, crossings and nestings for 
any $X \in \mathcal{M}_N$ must add up to the total number of pairwise chord
configurations, i.e., $N(N-1)/2$.
One of the earliest results~\cite{Errera,Jaglom} in the enumeration of chord 
diagrams is that the number of matchings in $\mathcal{M}_N$ with {\em no} 
crossings is given by the $N^{\rm th}$ Catalan number $C_N = \frac{1}{N+1}\binom{2N}{N}$,
which appears in many combinatorial problems (see, e.g.,~\cite{Stanley}).
Similarly, the number of diagrams in $\mathcal{M}_N$ with {\em no} nestings is 
also given by $C_N$.
The problem of counting the elements of $\mathcal{M}_N$ according to the
number of pairwise crossings of chords was considered by Touchard~\cite{Touchard},
who gave an implicit formula for the enumerating generating function in terms of
continued fractions. Subsequently, Riordan~\cite{Riordan} derived a remarkable 
explicit formula for the generating function based on Touchard's work.
If $cr(X)$ denotes the number of crossings of the element 
$X \in \mathcal{M}_N$, then the generating function by the number of crossings is 
defined via the polynomial
$$ F_N(q) := \sum_{X \in \mathcal{M}_N}\!\!q^{cr(X)}\,, \qquad 
0 \leq cr(X) \leq \half N(N-1) \,,$$
in the variable $q$ with positive integer coefficients.
The Touchard-Riordan formula for $F_N(q)$ is 
\begin{equation}
F_N(q) = \frac{1}{(1-q)^N}\sum_{n=0}^N\!\! (-1)^n\left[\binom{2N}{N-n}- 
\binom{2N}{N-n-1}\right]q^{n(n+1)/2}\,.
\label{e:TR}
\end{equation}
The first few polynomials are
$$ F_1(q) = 1, \quad F_2(q) = q+2, \quad F_3(q) = q^3 + 3q^2+6q+5 \,,$$  
and it easily follows from Eq.~\eqref{e:TR} that the number of non-crossing diagrams
is given by 
$$ F_N(0) = \binom{2N}{N}- \binom{2N}{N-1} = \frac{1}{N+1}\binom{2N}{N} \,,$$
which is the Catalan number $C_N$ mentioned earlier.
However, the Touchard-Riordan formula is somewhat mysterious, in  
that it is not obvious from Eq.~\eqref{e:TR} that $F_N(q)$ is in fact a polynomial 
in $q$ of degree $N(N-1)/2$ as implied by its combinatorial origin, or that 
$F_N(1) = |\mathcal{M}_N| = (2N-1)!!$. These assertions follow only after 
detailed analysis of Eq.~\eqref{e:TR}~\cite{Riordan} (see also \cite{Fla2000}).
A purely combinatorial proof of the Touchard-Riordan formula also appeared
in Ref.~\cite{Penaud} (see also \cite{Kasraoui}), and its relation to $q$-Hermite
polynomials was investigated in ref.~\cite{EJC87}.

%%%%%%%%%%%%%%%%%%%%  tau-function  %%%%%%%%%%%

\subsection{The $\tau$-function and line solitons of KPII}
It is well known (see e.g. \cite{Sato,Hirota})
that the solution $u(x,y,t)$ of the KPII equation is given in terms of
the $\tau$-function $\tau(x,y,t)$ as
\begin{equation}
u(x,y,t)= 2\frac{\partial^2}{\partial x^2}\ln\tau(x,y,t)\,.
\label{e:u}
\end{equation}
We consider the class of solutions whose $\tau$-functions are given by the Wronskian 
determinant form, i.e.,
\begin{equation}
\tau(x,y,t)= \Wr(f_1,\dots,f_N)= \det\begin{pmatrix}
  f_1 &f_2 &\dots &f_N\\
  f_1' &f_2' &\dots &f_N'\\
  \vdots &\vdots & &\vdots\\
  f_1^{(N-1)} &f_2^{(N-1)} &&f_N^{(N-1)}
  \end{pmatrix}\,.
\label{e:tau}
\end{equation}
with $f_n^{(j)}= \partial^j f_n/\partial x^j$, and
where the functions $\{f_n\}_{n=1}^N$
form a set of linearly independent solutions of the linear system
\[
\frac{\partial f}{\partial y}=\frac{\partial ^2f}{\partial x^2}\,,\qquad 
\frac{\partial f}{\partial t}=\frac{\partial ^3f}{\partial x^3}\,.
\]
The soliton solutions of KPII can be constructed from Eq.~\eqref{e:tau} by
choosing a finite dimensional solution for each function $f_n(x,y,t)$, namely,
\begin{equation}
f_n(x,y,t)= \sum_{m=1}^{M} a_{nm}\,e^{\theta_m}\,, \quad
n = 1,2, \ldots, N \,,
\label{e:f}
\end{equation}
where $\theta_m(x,y,t) = k_mx+k_m^2y+k_m^3t+\theta^{0}_{m}, \, m=1,\ldots,M$, 
are phases with real distinct parameters $k_1, k_2 \ldots k_M$,
and real constants $\theta^{0}_{1},\dots,\theta^{0}_{M}$. 
The constant coefficients $a_{nm}$ define the $N \times M$
\textit{coefficient matrix} $A:= (a_{nm})$.
The simplest example is the $1$-soliton solution with $N=1, M=2$, 
and $\tau=f_1=e^{\theta_1}+e^{\theta_2}$, for which   
\begin{equation*}
u(x,y,t)=2\frac{\partial^2}{\partial x^2}\ln\tau=
\frac{1}{2}(k_1-k_2)^2{\rm sech}^2\frac{1}{2}(\theta_1-\theta_2)\,.
%\label{e:onesoliton}
\end{equation*}
This $1$-soliton solution describes a plane traveling 
wave-form with constant amplitude $(k_1-k_2)^2/2$. For fixed $t$, the wave-form
is localized in the $(x,y)$-plane along the 
line $L: \theta_1=\theta_2$ whose normal has the slope $c= k_1+k_2$.
The solution is characterized by two physical parameters, namely,
the \textit{soliton amplitude parameter}~$a=|k_1-k_2|$ and the 
\textit{soliton direction parameter} $c=k_1+k_2$. 

In the general case, substitution of  
Eq.~\eqref{e:f} into the Wronskian of Eq.~\eqref{e:tau}, and subsequent
development of the resulting determinant via Binet-Cauchy formula,
yields the following explicit form of the $\tau$-function:
\begin{equation}
\tau(x,y,t)= \hspace{-0.1 in} \sum_{1\le m_1<\dots<m_N\le M} 
\hspace{-0.2 in}  A(m_1,\dots,m_N) \,\,
\exp[\,\theta(m_1,\dots,m_N) \,]
\!\!\prod_{1\le s < r\le N}(k_{m_{r}}-k_{m_{s}}) \,,
\label{e:tauexp}
\end{equation}
where $A(m_1,\dots,m_N)$ is the $N\times N$ maximal minor of $A$
obtained from the columns $1\le m_1<\dots<m_N\le M$,
and $\theta(m_1,\dots,m_N) := \theta_{m_1}+\ldots+\theta_{m_N}$
is a phase combination of $N$ (out of $M$) distinct phases.
Note that the transformation, $G: A \rightarrow A':=G\,A, \,\, G\in \mathrm{GL}(N,\Real)$,
amounts to an overall rescaling of the minors $A(m_1,\dots,m_N)$, and hence,
of the $\tau$-function in Eq.~\eqref{e:tauexp}; i.e., $\tau\to \tau'=\det(G)\,\tau$.
Since such a rescaling leaves the
solution $u(x,y,t)$ in Eq. \eqref{e:u} invariant, it is possible to reduce the
coefficient matrix $A$ to reduced row-echelon form (RREF) by Gaussian elimination.
Throughout the rest of this article, the coefficient matrix~$A$ will be assumed
to be in RREF.

The solutions $u(x,y,t)$ resulting from the $\tau$-function in Eq.~\eqref{e:tauexp} 
are singular for arbitrary choices of the parameters $\{k_n\}_{n=1}^M$ and the matrix $A$.
To avoid such singularities, which correspond to the zero-locus of the $\tau$-function,
one needs to impose certain positivity conditions.
\begin{condition}~(Positive definiteness of $\tau$).
\label{c:positive}
\begin{enumerate}
\item[(a)]\, The phase parameters are distinct, and are ordered 
as $k_1<k_2<\ldots<k_M$.
\item [(b)]\, The $N \times M$ coefficient matrix $A$ satisfies
$\rank(A) = N$, and $M > N$. 
\item[(c)]\, All non-zero $N \times N$ minors of $A$ are positive.
\end{enumerate}
\end{condition}
\begin{remark}
The matrices satisfying Condition \ref{c:positive}(c) above, are called
totally non-negative (TNN) matrices. The classification of the $(N_-,N_+)$-soliton
solutions is thus given by the classification of the $N\times M$ 
TNN matrices $A$ in RREF.
From a more geometric perspective, each TNN matrix
parametrizes a unique cell in the TNN Grassmannian
$Gr^+(N,M)$ (see e.g. \cite{postnikov:06}), and the classification of the soliton solutions
corresponds to a further refinement of the Schubert decomposition of $Gr(N,M)$ into
TNN Grassmann cells (see \cite{Kodama} for the case $M=2N$).
The refinement is given by a classification of the coefficient matrix $A$ whose
$N \times N$ minors $A(m_1,\ldots,m_N)$ represent the Pl\"ucker coordinates of $Gr(N,M)$.
It should be noted that each $(N_-,N_+)$-soliton solution corresponding to a TNN matrix
can be  parametrized by a chord diagram \cite{mpag,Corteel}.
The geometric structure of this classification will be discussed in a future 
communication~\cite{CK}.
\end{remark}
In Condition \ref{c:positive}, the distinctness assumption on the set of phase 
parameters ensures that
the set $\{e^{\theta_m}\}_{m=1}^M$ is linearly independent, while
$\rank(A) = N$ implies that the set of functions $\{f_n\}_{n=1}^N$
is linearly independent. Also when $M=N$, the sum in
Eq.~\eqref{e:tauexp} reduces to a single exponential term with a
phase combination that is linear in $x$. The logarithm of such a $\tau$-function
is annihilated by the second derivative in Eq.~\eqref{e:u}, leading 
to the trivial solution $u(x,y,t) =0$.
Thus the condition $N < M$ guarantees non-trivial solutions of KPII.
Condition~\ref{c:positive}(c) together with the ordering
$k_1<k_2<\ldots<k_M$ make the sum in Eq.~\eqref{e:tauexp} totally
positive. As a result, $\tau(x,y,t)$ is a positive function on $\Real^3$,
and the resulting solution $u(x,y,t)$ of KPII is non-singular, bounded and positive 
definite. Furthermore, the asymptotic analysis of the $\tau$-function in 
Eq.~\eqref{e:tauexp} reveals that for any given value of~$t$, there exist a set 
of lines given by $\{L_{ij}:\,\theta_i=\theta_j,\,\, i < j\}$ in the $(x,y)$-plane,
such that
\begin{equation}
u(x,y,t) \sim \half(k_j-k_i)^2\sech^2\half(\theta_j-\theta_i+\delta_{ij}) \,,
\label{e:uasymp}
\end{equation}
along each $L_{ij}$ either as $y \to \infty$, or as $y \to -\infty$. 
Equation \eqref{e:uasymp}, which has the same form as the $1$-soliton solution
defines an {\em asymptotic line soliton} along $L_{ij}$ associated with 
the solution $u(x,y,t)$. Each line soliton which is parallel to the line $L_{ij}$ has 
the parameters $a_{ij}=|k_i-k_j|$ for the amplitude
and $c_{ij}= k_i+k_j$ for the direction normal to the line $L_{ij}$.
Hence, we denote each line soliton by the index pairing ${\bf p}=[i,j]$ labeling
the line $L_{ij}$. Note from Eq.~\eqref{e:f} that the indices $i, j$ labeling
the phases $\theta_i,\, \theta_j$ in Eq. \eqref{e:uasymp} also label a pair of distinct 
columns of the coefficient matrix $A$. Due to this connection, it turns out that 
the pairing: ${\bf p}=[i,\,j],\,\, 1\leq i < j \leq M$, of the line solitons can be 
determined from the structure of the coefficient matrix $A$ which is in RREF, and 
satisfies the following irreducibility condition.
\begin{condition} (Irreducibility)~
\label{c:irreducible}
\begin{itemize}
\item[(a)] Each column of $A$ contains at least one nonzero element.
\item[(b)] Each row of $A$ contains at least one nonzero element in addition to the pivot.
\end{itemize}
\end{condition}
Recall that, for an $N\times M$ matrix in RREF,
the leftmost non-vanishing entry in each nonzero row is called a pivot,
which is normalized to unity.
The index pairs $[i,j]$ of the asymptotic line solitons shown in Eq.~\eqref{e:uasymp} 
are then given by the following technical result, which is proved in Ref.~\cite{BC}.
\begin{proposition}
\label{P:pairing}
Let the sub-matrices $X[ij]$ and $Y[ij]$ of $A$ be defined 
in terms of their column indices as
\begin{equation*}
X[ij] := \left[ 1,2,\ldots, i-1, j+1, \ldots, M \right] \qquad
Y[ij] := \left[i+1, \ldots j-1 \right] \,.
%\label{e:XY}
\end{equation*}
Then, necessary and sufficient conditions for an index pair $[i,j]$
to specify an asymptotic line soliton, as in Eq.~\eqref{e:uasymp}, are the following 
rank conditions.
\begin{enumerate}
\item[(i)] Each line soliton as $y\to\infty$ is labeled by
a unique index pair $[e_n,j_n]$ with $e_n < j_n$, where
$\{e_n\}_{n=1}^N$ label the pivot columns of $A$.
Moreover, if $\rank(X[e_nj_n])=:r_n$, then $r_n \le N-1$ and
$$ \rank(X[e_n \, j_n]|e_n) \,\, = \,\,\rank(X[e_n \, j_n]|j_n) \,\,
= \,\, \rank(X[e_n\, j_n]|e_n, j_n) \,\,= \,\,r_n+1 \,.$$
\item[(ii)] Each line soliton as $y\to -\infty$ is labeled
by a unique index pair $[i_n,g_n]$ with $i_n < g_n$, where
$\{g_n\}_{n=1}^{M-N}$ label the non-pivot columns of $A$.
Moreover, if $\rank(Y[i_ng_n]) =: s_n$, then $ s_n \le N-1$ and
$$ \rank(Y[i_n \, g_n]|i_n) \,\, = \,\, \rank(Y[i_n \, g_n]|g_n)
\,\, = \,\,\rank(Y[i_n \, g_n]|i_n, g_n)\,\,= \,\,s_n+1 \,.$$
\end{enumerate}
Here $(Z|m,n)$ denotes the sub-matrix~$Z$ of $A$ augmented by the
columns $m$ and $n$ of $A$.
\end{proposition}
It should be clear from the above result that the $(N_-,N_+)$-soliton
solution of KPII generated from the $\tau$-function
in Eq.~\eqref{e:tauexp} has exactly $N_+= N$ asymptotic line-solitons
as $y\to\infty$ and $N_-=M-N$ asymptotic line-solitons as $y\to-\infty$.
From the $\tau$-function data consisting of $M$ distinct phase parameters
$k_1,\ldots,k_M$ and a matrix $A$ satisfying Condition \ref{c:irreducible},
Proposition \ref{P:pairing} provides an explicit
way to identify all the asymptotic line solitons of the corresponding
solution of the KPII equation. We illustrate this method with an example.
\begin{example}
Consider the solution $u(x,y,t)$ generated by the $\tau$-function 
of Eq.~\eqref{e:tau} in the case $N=2$ and $M=4$, with 4 real parameters 
$k_1 < k_2 < k_3 < k_4$, and
$$
f_1 = e^{\theta_1}-e^{\theta_4}, \quad
f_2 = e^{\theta_2}+e^{\theta_3}, \qquad \qquad
A= \begin{pmatrix}
1 &0 &0 &\!-1 \\ 0 &1 &1 &0
\end{pmatrix} \,.
$$
The pivot columns of~$A$ are labeled by the indices $\{e_1,e_2\}=\{1,2\}$,
and the non-pivot columns by the indices $\{g_1,g_2\}=\{3,4\}$.
According to Proposition~\ref{P:pairing}, the number of asymptotic line
solitons is $N_+=N_-=2$. They are identified by the index pairs $[1,j_1],\, [2,j_2]$
as $y\to \infty$, for some $j_1>1$ and $j_2>2$; and by the index pairs $[i_1,3]\,,[i_2,4]$
as $y\to -\infty$, for some $i_1<3$ and $i_2<4$.
We first determine the asymptotic line-solitons as $y\to\infty$
using the rank conditions prescribed in Proposition~\ref{P:pairing}(i).
For the first pivot column $e_1=1$; starting from $j=2$ and then repeatedly
incrementing the value of $j$ by unity, we check the rank of each sub-matrix $X[1j]$.
Proceeding in this way, we find that the rank conditions are satisfied
{\em only} when $j=4$: $X[14]= \emptyset$.
So, $\rank(X[14])=0 <N-1$. Moreover, $\rank(X[14]|1)= \rank(X[14]|4)=
\rank(X[14]|1,4) =1$ since columns 1 and 4 are parallel.
Thus, the first asymptotic line soliton
as $y\to\infty$ is identified by the index pair $[1,4]$.
For $e_2=2$, proceeding in a similar manner we
find that $j=3$ does satisfy the rank conditions since
$X[23] = \bigl(\begin{smallmatrix}1&-1\\0&0\end{smallmatrix}\bigr)$ is of
rank~$1=N-1$, and $\rank(X[23]|2)= \rank(X[23]|3)= \rank(X[23]|2,3) =2$.
Therefore, the asymptotic line solitons as $y\to\infty$ are identified with 
the index pairs $[1,4]$ and $[2,3]$.

We next consider the asymptotics for $y\to-\infty$.
Starting with the non-pivot column $g_1=3$, we apply the rank conditions
in Proposition~\ref{P:pairing}(ii) to the column $i=2$.
Then, we have $Y[23]=\emptyset$, and
$\rank(Y[23]|2)= \rank(Y[23]|3)= \rank(Y[23]|2,3)=1$.
Hence, the pair $[2,3]$ identifies an asymptotic line-soliton
as $y\to-\infty$. For $g_2=4$, we consider $i=1,2,3$ and find that
the rank conditions are satisfied only for $i=1$. In this case,
$Y[14]= \bigl(\begin{smallmatrix}0&0\\1&1\end{smallmatrix}\bigr)$,
so $\rank(Y[14])= 1=N-1$ and
$\rank(Y[14]|1)= \rank(Y[14]|4)= \rank(Y[14]|1,4)=2$.
Thus, the index pair $[1,4]$ identifies the other
asymptotic line-soliton as $y\to-\infty$.
In summary, both pairs of asymptotic line solitons as $y \to \pm \infty$ are labeled
by the index pairs $[1,4]$ and $[2,3]$. This is an example of a P-type $2$-soliton solution
(see Section~\ref{s:elastic}), and is shown in Fig.~\ref{f:2sol}.
\label{E:Ptype}
\end{example}
It should be emphasized that in general $N_- \neq N_+$, and that even in the case
$N_- = N_+$, the line solitons as $y \to \infty$ are
in general distinct from the line solitons as $y \to -\infty$ in both
amplitude and direction. In this article, we restrict our discussions
primarily to the $N$-soliton subclass of the line-soliton solutions of KPII.
\begin{definition}
Let $S_+ := \{[e_n,j_n]\}_{n=1}^N$ and $S_- := \{[i_n,g_n]\}_{n=1}^{M-N}$ denote
the index sets identifying the line solitons as $y \to \infty$ and
as $y \to -\infty$, respectively, according to  Proposition \ref{P:pairing}.
Then two $(N_-,N_+)$-soliton solutions of KPII are
said to be in the same equivalence class if their asymptotic line-solitons are labeled
by identical sets $S_\pm$ of index pairs, where $|S_+| := N_+=N$ and
$|S_-| := N_- = M-N$.
\label{D:equiv}
\end{definition}
The set $S_+ \cup S_-$ of unique index pairings in Definition \ref{D:equiv}
has a combinatorial interpretation.
Let $[M] := \{1,2,\ldots,M\}$ be the integer set with the pivot and non-pivot
indices $\{e_1,\ldots,e_N\} \cup \{g_1,\ldots,g_{M-N}\}$ forming a disjoint partition
of $[M]$. Define the pairing map $\pi: [M] \to [M]$
according to Proposition \ref{P:pairing}(i) \& (ii) as
\begin{equation}
\pi(e_n) = j_n\,, \,\, n=1,2,\ldots,N\,, \qquad
\pi(g_n) = i_n\,, \,\, n=1,2,\ldots,M-N\,.
\label{e:pair}
\end{equation}
Then $\pi: [M] \to [M]$ is a bijection, i.e., $\pi \in \mathcal{S}_M$, the permutation
group of $[M]$~\cite{mpag}. In addition, Proposition \ref{P:pairing} implies the following.
\begin{proposition}
\label{P:derangement}
The pairing map $\pi$ defined by Eq.~\eqref{e:pair} is a derangement of $[M]$
with $N$ excedances, which are given by the pivot indices $\{e_1,\ldots,e_N\}$
of the coefficient matrix $A$ in RREF.
\end{proposition}
Recall that a permutation $\pi$ with no fixed
point is called a {\em derangement}, and an element $l \in [M]$ is called an
{\em excedance} of $\pi$ if $\pi(l) > l$.

Each equivalence class of $(N_-,N_+)$-soliton solutions of KPII is uniquely
determined by a derangement $\pi$, as in Proposition \ref{P:derangement}. These
derangements also give a unique parametrization of a TNN Grassmann cell 
(see Remark 2.3) whose associated TNN matrix $A$ satisfies the irreducibility
Condition~\ref{c:irreducible}.
Recall that in Example \ref{E:Ptype}, both sets of line solitons as 
$y \to \pm \infty$ are given by $[1,4],\,[2,3]$, where 1,2 are the pivot indices
and 3,4 are the non-pivot indices of the associated coefficient matrix $A$.
In this case, the pairing map is a derangement of the set $[4]$, and is given by
\[
 \pi= \begin{pmatrix}
1 &2 &3 &4\\ 4 &3 &2&1
\end{pmatrix} \in \mathcal{S}_4\,,
\]
in the bi-word notation of permutations in $\mathcal{S}_4$. Note that the excedance set 
of $\pi$ is $\{1,2\}$. In addition, $\pi$ is also an involution of $\mathcal{S}_4$,
i.e., $\pi^{-1} = \pi$. Since the set of all involutions of $\mathcal{S}_{2N}$ is isomorphic
to the set of perfect matchings $\mathcal{M}_N$ introduced in 
Section~\ref{s:background}.1, the involutions can be also
represented by the chord diagrams representing the elements of $\mathcal{M}_N$.
In particular, the chord diagram for the involution $\pi \in \mathcal{S}_4$ 
given above depicts a nesting of the chords ${\bf p}_1 = [1,4]$ and ${\bf p}_2 = [2,3]$
as shown below in Fig.~\ref{f:2sol}. We remark that it is possible to represent
derangements that are not involutions by linear chord diagrams, with directed chords both
above and below the line. These diagrams have been used to study the more general 
$(N_-,N_+)$-soliton solutions of KPII in Ref.~\cite{mpag}. But here we focus our attention 
to the $N$-soliton solutions, which will be our next topic of discussion.

%%%%%%%%%%%%%%%%%%%%%%%%%%%%%%%%%%%%%%%%%%%%%%%%%%%%%%%%%%%%%%%%%%%%%%%%%%%%%%%%%%%%%%
\section{$N$-soliton solutions}
\label{s:elastic}
When $M=2N$, it follows from Proposition \ref{P:pairing} that $N_-=N_+ = N$.
If in addition, we consider $S_- = S_+$ in Definition~\ref{D:equiv},
then we recover the interesting subclass of the $(N,N)$-soliton solutions mentioned
in Section~\ref{s:introduction}, called $N$-soliton solutions, which
are characterized by identical sets of asymptotic
line-solitons as $|y| \to \infty$.
Then the main features of the $N$-soliton solutions follow primarily from
our discussion in Section~\ref{s:background}, in particular
from Propositions~\ref{P:pairing} and ~\ref{P:derangement}.
These are listed below.
\begin{property} $N$-soliton solutions have the following properties.
\begin{enumerate}
\item[(i)] The $\tau$-function of an $N$-soliton solution is expressed
in terms of $2N$ distinct phase parameters and an $N \times 2N$
coefficient matrix $A$ which satisfies Conditions~\ref{c:positive} 
and \ref{c:irreducible}. In addition, the
$N \times N$ minors of $A$ satisfy the duality conditions~\cite{mpag, Kodama}:
\begin{equation*}
A(m_1,\ldots,m_N) = 0 \qquad \Longleftrightarrow \qquad A(l_1,\ldots,l_N) = 0\,,
\label{e:dualminor}
\end{equation*}
where the indices $\{m_1,\ldots,m_N\}$ and $\{l_1,\ldots,l_N\}$ form a disjoint
partition of integers $\{1,2,\ldots, 2N\}$. That is, the phase combination
$\theta(m_1,\ldots,m_N)$ is present in the $\tau$-function of Eq.~\eqref{e:tauexp}
 if and only if $\theta(l_1,\ldots,l_N)$ is.
\item[(ii)] Each $N$-soliton solution has exactly $N$
asymptotic line solitons as $y \to \pm \infty$ identified by the
same index pairs $[e_n,g_n]$ with $e_n < g_n, \, n=1,\ldots, N$.
The sets $\{e_1, \ldots, e_N\}$ and $\{g_1, \ldots, g_N\}$ label
respectively the pivot and non-pivot columns of the coefficient matrix
$A$. Hence, they form a disjoint partition of the integer set $[2N]$.
\item[(iii)] The amplitude and direction parameters of the $n^{\mathrm {th}}$ asymptotic line
soliton $[e_n,g_n]$ are the same as $y \to \pm \infty$, and
are given in terms of the phase parameters as 
\begin{equation*}
a_n = k_{g_n} - k_{e_n}\,, \qquad \qquad c_n = k_{g_n} + k_{e_n}\,.
\label{e:solitonparameters}
\end{equation*}
\item[(iv)] The pairing map associated to an $N$-soliton solution, namely
$\pi(e_n) = g_n,\, \pi(g_n) = e_n, \,\, n=1,2,\ldots,N$, corresponds to a partition
of the integer set $[2N]$ into $N$ distinct pairs of integers $(e_n,g_n)$, as
in Section~\ref{s:background}.1. Each such map is an involution in $\mathcal{S}_{2N}$ 
with no fixed points, a member of 
$$\mathcal{I}_{2N} = \{\pi \in \mathcal{S}_{2N}:\,\pi^{-1}=\pi \, 
\mbox{and} \, \pi(i) \neq i,\, \forall\, i \in [2N]\}\,.$$
Such permutations can be expressed as products of $N$ disjoint $2$-cycles,
and their chord diagrams are identical to those of the perfect matchings 
$\mathcal{M}_N$ (see Fig.~\ref{f:chord1a}). The total number of such involutions
is given by $|I_{2N}| = |\mathcal{M}_N| = (2N-1)!!$.
Hence, there are $(2N-1)!!$ distinct equivalence classes of $N$-soliton solutions.
\end{enumerate}
\label{p:Nsoliton}
\end{property}
\subsection{Equivalence classes of 2-soliton solutions}
\begin{figure}[h]
\centering
\includegraphics[scale=0.58]{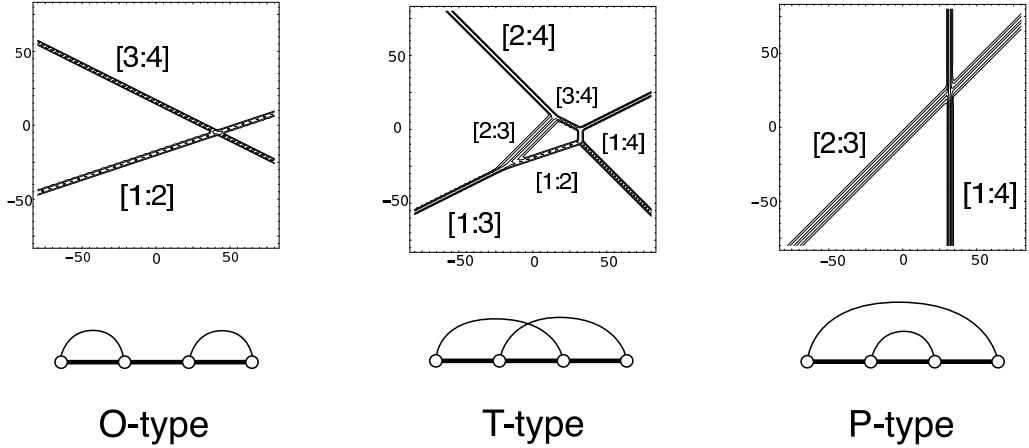}
\caption{Three different two-soliton solutions of KPII with the same
phase parameters~$(k_1,\dots,k_4)=(-2,-\frac12,0,1)$,
illustrating the three $2$-soliton equivalence classes:
O-type,~T-type and ~P-type 2-soliton solutions.}
\label{f:2sol}
\end{figure}

When $N=2$, there are three types of 2-soliton solutions
referred to as the O-, T- and P-types (following the
terminology introduced in Ref.~\cite{Kodama}). They are
identified by the canonical coefficient matrices associated with
$\tau$-functions, namely
\begin{gather}
A_{\mathrm{O}}=
\begin{pmatrix}1 &1 &0 &0\\ 0 &0 &1 &1\end{pmatrix},
\qquad
A_{\mathrm{T}}=
\begin{pmatrix}1 &0 &-1 &-1\\ 0 &1 &x_1 &x_2\end{pmatrix},
\qquad
A_{\mathrm{P}}=
\begin{pmatrix}1 &0 &0 &-1\\ 0 &1 &1 &0\end{pmatrix},
\label{e:Acanonical}
\end{gather}
with $x_1>x_2>0$ in~$A_{\mathrm{T}}$.
By applying the rank conditions of Proposition~\ref{P:pairing} to the
above coefficient matrices, it is easily verified that the O-type 2-solitons
have asymptotic line-solitons [1,2] and [3,4]; the T-type resonant 2-solitons
have asymptotic line-solitons [1,3] and [2,4]; and the P-type 2-solitons
have asymptotic line-solitons [1,4] and [2,3]. These are shown in Fig.~\ref{f:2sol}.
Notice that each of the O- and P-type solitons interact via an X-junction. After 
interaction, each line soliton undergoes a position shift in the $xy$-plane. However,
it can be shown that the position shifts for the O-type solitons are {\em opposite}
in sign to that of the P-type solitons~\cite{mpag}.
On the other hand, the T-type solitons interact via four
Y-junctions, connecting the four asymptotic line-solitons to four
intermediate segments. Each of these intermediate segments is also a line soliton.
For example, in the T-type solitons in Fig.~\ref{f:2sol} the asymptotic line
soliton $[1,3]$ (as $y \to -\infty$) forms the intermediate line-solitons
$[1,2]$ and $[2,3]$ at the bottom left Y-junction. The line-soliton $[2,3]$
connects with the asymptotic line-soliton $[2,4]$ (as $y \to \infty$)
and the line-soliton $[1,2]$ connects with the asymptotic line
soliton $[2,4]$ (as $y \to -\infty$). Similarly, the asymptotic line
soliton $[1,3]$ (as $y \to \infty$) forms the intermediate line-solitons
$[1,4]$ and $[3,4]$ at the top right Y-junction. The line-soliton $[3,4]$
connects with the asymptotic line-soliton $[2,4]$ (as $y \to \infty$)
and the line-soliton $[1,4]$ connects with the asymptotic line
soliton $[2,4]$ (as $y \to -\infty$).
Fig.~\ref{f:2sol} also shows the chord diagrams for the corresponding pairing maps,
which are involutions of the permutation group ${\mathcal{S}}_4$. 
They correspond to the disjoint partitions
of $[4]$ into 2 pairs. In cycle notation, these involutions are given by
$\pi_{\mathrm{O}} = (12)(34),\,\pi_{\mathrm{T}}=(13)(24)$, and $\pi_{\mathrm{P}}=(14)(23)$,
for the O-, T- and P-type $2$-soliton equivalence classes. According to the
Definition~\ref{d:config}, the chord diagram for $\pi_{\mathrm{O}}$ forms an alignment; 
whereas the diagram for $\pi_{\mathrm{T}}$ has a crossing between the chords 
corresponding to the line solitons $[1,3]$ and $[2,4]$; and the diagram for
$\pi_{\mathrm{P}}$ is a nesting. 

An important distinction among the three types of 2-soliton solutions
is that they belong to different regions of the soliton parameter space.
Suppose $(a_1,c_1)$ and $(a_2,c_2)$ are the soliton parameters of the
asymptotic line-solitons of each type, with the same set of distinct phase
parameters. Since the phase parameters are ordered:~$k_1<\dots<k_4$,
the soliton parameters satisfy the following relations,
which can be easily verified using Eqs.~\eqref{e:solitonparameters}.
\begin{enumerate}
%\advance\itemsep-5pt
\item
For O-type 2-soliton solutions, $c_2>c_1$ and $c_2-c_1>a_1+a_2$.
\item
For T-type 2-soliton solutions, $c_2>c_1$, and $|a_1-a_2|<c_2-c_1<a_1+a_2$.
\item
For P-type 2-soliton solutions, $a_2>a_1$ and  $|c_2-c_1|<a_2-a_1$.
\item
$(c_2-c_1)_\mathrm{O}>(c_2-c_1)_\mathrm{T}>|c_2-c_1|_\mathrm{P}$\,, \quad
$(a_1+a_2)_\mathrm{O}<(a_1+a_2)_\mathrm{T}=(a_1+a_2)_\mathrm{P}$\,, and \\
$|a_2-a_1|_\mathrm{O}=|a_2-a_1|_\mathrm{T}<(a_2-a_1)_\mathrm{P}$\,.
\end{enumerate}
Note that for O- and T-type solutions the soliton directions are ordered,
while for P-type solutions the amplitudes are ordered. {\em Any} choice
of the soliton parameters $a_1,c_1;a_2,c_2$ with $a_1, a_2 > 0$
would lead to one of the three types of 2-soliton solutions, provided that
$\{c_1 \pm a_1,\, c_2 \pm a_2\}$ are distinct real numbers.
Thus, the three types of 2-soliton solutions partition the soliton parameter
space into disjoint sectors, bounded by the hyperplanes
$|c_2-c_1|= a_1 + a_2$ and $|c_2-c_1|= |a_1 - a_2|$.
At each boundary between two sectors, two of the phase parameters coincide.
In such a situation, it can be shown (by taking suitable limits)
that the 2-soliton solution degenerates into a Y-junction~\cite{jphysa36p10519,Kodama}.

It should be clear from the above that the O-, T- and P-type $2$-soliton solutions 
exhibit distinct types of interaction patterns, and belong to different regions of the
soliton parameter space. For $N>2$, in addition to the 
non-resonant (O- and P-type) and fully resonant (T-type) solutions, a large family 
of partially resonant solutions exists. For example, when $N=3$, 
Property \ref{p:Nsoliton}(iv) implies that there are 15 distinct equivalent classes of 
$3$-soliton solutions (see Fig.\ref{f:N3chord}). Unlike the $N=2$ case above, 
it turns out to be a complicated task to classify the $N$-soliton solutions according 
to their coefficient matrices $A$, when $N > 2$. This task was recently carried out 
by the authors, and will be reported in a future publication~\cite{CK}.
Here we consider a more direct classification scheme for the $N$-soliton equivalence 
classes, by characterizing the pairwise interactions between the $N$ line solitons of 
O-, T- and P-type, much as in the $2$-soliton case.  
In other words, we represent the $N$-soliton solutions by the corresponding
involutions in $\mathcal{I}_{2N} \subset \mathcal{S}_{2N}$ 
(equivalently, the matchings in ${\mathcal M}_N$), and enumerate the solutions
according to the number of alignments, crossings and nestings of the
associated chord diagram. We describe this classification scheme below. 

\subsection{Combinatorics of $N$-soliton solutions}
\begin{figure}[t!]
\centering
\includegraphics[scale=0.55]{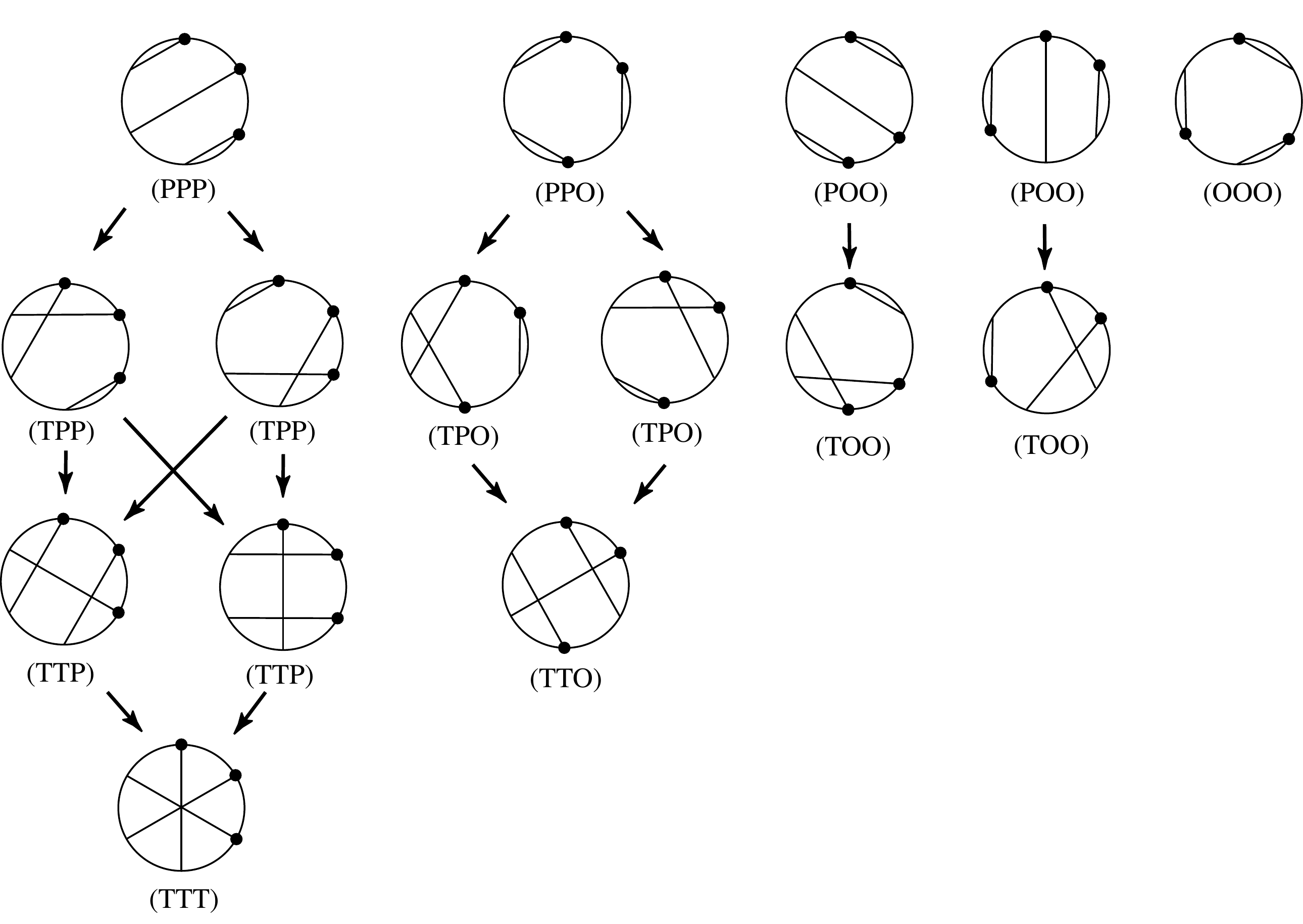}
\caption{The closed chord diagrams for $3$-soliton solutions. The dots  
indicate the pivots $(e_1,e_2,e_3)$, and the ordered letters below each diagram 
indicate the type of interactions in 
$({\bf p}_1{\bf p}_2,{\bf p}_2{\bf p}_3,{\bf p}_3{\bf p}_1)$ with
the soliton pairing ${\bf p}_n=[e_n,g_n]$. The number of the diagrams having the 
same number of crossings comes from the generating function $F_3(q)=q^3+3q^2+6q +5$.
E.g., 5, the Catalan number $C_3=F_3(0)$, counts the diagrams in the first row.}
\label{f:N3chord}
\end{figure}

Throughout this subsection, we associate an $N$-soliton equivalence class
defined by the set $S=\{[e_n,g_n]\}_{n=1}^N$ of asymptotic line solitons 
(see Definition~\ref{D:equiv}) with the partition 
$X = \{ {\bf p}_1, {\bf p}_2, \ldots , {\bf p}_N\} \in \mathcal{M}_N$, where 
${\bf p}_n := [e_n,g_n]$. Recall from Property~\ref{p:Nsoliton}(ii) that the integer 
set $[2N]$ is a disjoint union of the index sets $E := \{e_1,\ldots,e_N\}$ and 
$G:=\{g_1,\ldots,g_N\}$, with the following orderings among the indices:

(i)\,  $1=e_1 < e_2 < \ldots < e_N < 2N$, 

(ii) $e_n < g_n$ for all $n=1,2,\ldots,N$. \\
An immediate consequence of the above orderings is that
\begin{equation}
n \leq e_n \leq 2n-1\,, \qquad n=1,\ldots, N\,,
\label{e:pivot}
\end{equation}
since there are at least $n-1$ indices to the left of $e_n$, namely 
$e_1,e_2, \ldots, e_{n-1}$; and at least $2N-2n+1$ indices to the right of
$e_n$, namely $g_n, e_r, g_r, \,\, r>n$. 
The $N$-soliton classification scheme is obtained by considering various statistics 
over the possible chord configurations for the chord diagrams of $\mathcal{M}_N$.
For this purpose, using Definition~\ref{d:config} we introduce the following sets, 
which record the total number of alignments, crossings and nestings 
for a given chord in any chord diagram of $\mathcal{M}_N$.
\begin{definition}
Let ${\bf p}_n = [e_n,g_n]$ be a given chord of a partition $X \in \mathcal{M}_N$,
and let $B_n := \{{\bf p}_r=[e_r,g_r]\,:\,r < n\}$ be the subset of chords originating
from the left of ${\bf p}_n$ in the linear chord diagram of $X$.
\vspace{-0.1 in}
\begin{enumerate}
\item[(a)]\, The set $O_n$ of alignments with the chord ${\bf p}_n$ forming O-type
configurations, and the alignment number $al(X)$, are defined by
\vspace{-0.2 in}
$$ O_n := \{{\bf p}_r=[e_r,g_r] \in B_n\,:\, g_r < e_n\}\,, \qquad
al(X) := \sum_{n=1}^N |O_n|\,.$$
\vspace{-0.2 in}
\item[(b)]\, The set $T_n$ of crossings with the chord ${\bf p}_n$ forming T-type
configurations, and the crossing number $cr(X)$, are defined by 
\vspace{-0.2 in}
$$ T_n := \{{\bf p}_r=[e_r,g_r] \in B_n\,:\, e_n < g_r < g_n\}\,, \qquad
cr(X) := \sum_{n=1}^N |T_n|\,.$$
\vspace{-0.2 in}
\item[(c)]\, The set $P_n$ of nestings with the chord ${\bf p}_n$ forming P-type
configurations, and the nesting number $ne(X)$, are defined by 
\vspace{-0.2 in}
$$ P_n := \{{\bf p}_r=[e_r,g_r] \in B_n\,:\, g_r > g_n\}\,, \qquad
ne(X) := \sum_{n=1}^N |P_n|\,.$$
\end{enumerate}
\label{d:OTP}
\end{definition}
\vspace{-0.1 in}
It follows from the above definitions that $B_n$ is the disjoint union of
the sets $O_n,\, T_n$ and $P_n$, so that $|O_n|+|T_n|+|P_n| = n-1$,
and $al(X) + cr(X) + ne(X) = N(N-1)/2$, which is a count of all possible pairwise
chord configurations in the partition $X$.
Note that for $O_n$, the indices $g_r$ lie in the intervals 
$(e_r,e_{r+1}),\,\, 1 \leq r < n$. Hence, $|O_n| = e_n-n$. So the number of 
crossings and nestings with the chord ${\bf p}_n$ sum to 
$|T_n|+|P_n| = (n-1)-(e_n-n) = 2n-e_n -1$, which depends {\em only
on the pivot index} $e_n \in E$. This observation leads to the following.
\begin{lemma}
If $\mathcal{M}(E) \subseteq \mathcal{M}_N$ denotes the set of all partitions
which have the same (pivot) index set $E$, then the number of partitions
of $\mathcal{M}(E)$ having $r$ crossings and $s$ nestings is the coefficient
of $p^sq^r$ in
$$ m_E(p,q) = \prod_{n=1}^N\,[2n-e_n]_{p,q}\,, \qquad 
[n]_{p,q} := \frac{p^n-q^n}{p-q} = \sum_{i+j = n-1}p^iq^j\,.$$
The degree of both $p$ and $q$ in $m_E(p,q)$ is $N^2 - (e_1+e_2+\ldots+e_N)$.
\label{L:crossing}
\end{lemma}
\begin{proof}
The distribution of crossings and nestings is the sum of $p^{ne(X)}q^{cr(X)}$
over all partitions $X \in \mathcal{M}(E)$. Using Definition~\ref{d:OTP}
for $cr(X)$ and $ne(X)$, this distribution can be expressed as
$$\sum_{X \in \mathcal{M}(E)}p^{(|P(1)|+\ldots+|P(N)|)}q^{(|T(1)|+\ldots+|T(N)|)} 
= \prod_{n=1}^N \sum_{l=0}^{2n-e_n-1}p^lq^{2n-e_n-1-l}\,, $$
after interchanging the sum and product, and using the fact that $|T_n|+|P_n| = 2n-e_n-1$
for $n =1,2,\ldots,N$. Since the second sum is precisely $[2n-e_n]_{p,q}$, the
formula for $m_E(p,q)$ follows. 
\end{proof}
It is easy to verify from the product formula that $m_E(p,q)$ is
symmetric in $p$ and $q$. Consequently, the number of diagrams with
$r$ crossings and $s$ nestings is the same as the the number of diagrams
with $s$ crossings and $r$ nestings~\cite{Kasraoui}. Note also that the enumerating 
polynomial for the crossings alone is given by $m_E(1,q)$; while  
$m_E(p,1)$ enumerates only the nestings for the chord diagrams of $\mathcal{M}(E)$.
In order to extend the results of Lemma~\ref{L:crossing},
to the entire set $\mathcal{M}_N$, one needs to sum $m_E(p,q)$ over all possible
choices of the integer set $E$, with $e_n \in E$ satisfying Eq.~\eqref{e:pivot}.
Using Lemma~\ref{L:crossing}, the expression for the required generating polynomial 
is given by
\begin{equation}
F_N(p,q) := \sum_{X \in \mathcal{M}_N}p^{ne(X)}q^{cr(X)} = \sum_{\{E\}}m_E(p,q)
= \sum_{\genfrac{}{}{0pt}{1}{1=e_1 < e_2 < \ldots <e_N,}{k\leq e_k\leq 2k-1}}\,
\prod_{n=1}^N [2n-e_n]_{p,q}\,. 
\label{e:Fpq}
\end{equation}
Since we have from Eq.~\eqref{e:pivot} that $e_n \geq n, \, n=1,2,\ldots,N$,
it follows from Lemma~\ref{L:crossing} that the degree of $p$ and $q$
in $F_N(p,q)$ is given by 
$$ ne(X)_{\mathrm max} = cr(X)_{\mathrm max} = N^2-(1+2+\ldots+N) = \frac{N(N-1)}{2} \,.$$
Furthermore, like $m_E(p,q)$, $F_N(p,q)$ is symmetric in $p$ and $q$, i.e.,
\begin{equation*}
F_N(p,q) = \sum_{r,s=0}^{N(N-1)/2}c_{rs}q^rp^s\,, \qquad c_{rs}=c_{sr}\,.
%\label{e:Fpq1}
\end{equation*}
Some interesting consequences of Eq.~\eqref{e:Fpq} for special cases of $F_N(p,q)$
are collected below.
\begin{corollary} The function $F_N(p,q)$ has the following properties:
\begin{itemize}
\item[(i)] $F_N(1,1) = |\mathcal{M}_N| = (2N-1)!!$. 
\item[(ii)] When $p=1, q=0$, the total number of non-crossing (i.e., only alignments and nestings)  
chord diagrams of $\mathcal{M}_N$ equals the $N^{\mathrm th}$ Catalan number~\cite{Errera}.
That is, $F(1,0) = C_N$ which also counts the total possible choices for the
ordered integer set $E$~\cite{mpag}. Similarly, $F_N(0,1) = C_N$ gives
the total number of non-nesting (i.e., only alignments and crossings) chord diagrams of
$\mathcal{M}_N$.
\item[(iii)] $F_N(1,q)=: F_N(q)$ gives the generating polynomial for the number
of crossings introduced in Section~\ref{s:background}.1, which is given explicitly
by the Touchard-Riordan formula Eq.~\eqref{e:TR}.
\end{itemize}
\label{c:specialpq}
\end{corollary}
The polynomials $F_N(p,q)$ can be determined from a generating function
$F(p,q,x)$ which is a formal power series, and has the following representation.
\begin{proposition}
The generating function for $F_N(p,q)$ is the Stieltjes-type continued fraction, namely
\[
F(p,q,x):= \sum_{N=0}^{\infty}F_N(p,q)x^N =\cfrac{1}{1-
                                            \cfrac{x\,[1]_{p,q}}{1-
                                             \cfrac{x\,[2]_{p,q}}{1-
                                              \cfrac{x\,[3]_{p,q}}{1-\cdots}}}} \,, 
\qquad F_0(p,q):=1\,.
\]
\label{P:gen}
\end{proposition}
\begin{proof}
First consider the set $E:=\{1=e_1<\cdots<e_N : e_k \leq 2k-1\}$.
Note that $E$ can be decomposed
into distinct subsets when $e_n=2n-1$. One has
$$ E = \bigcup_{n=0}^{N-1} \left(E_n \cup \hat{E}_n\right) \,,$$
where $E_n :=\{1=e_1<\cdots<e_n:\,e_k \leq 2k-1\}$ for $n \neq 0$ can be viewed
as the $n$-truncates of the original set $E$, $E_0 = \varnothing$, and
$$\hat{E}_n := \{2n+1=e_{n+1}<\cdots<e_N:\, 2n+k \leq e_{n+k} < 2(n+k)-1\} \,.$$
The set $\hat{E}_n$ can be re-expressed as
$$
E'_{N-n}= \{1=e'_1<\cdots<e'_{N-n}:\, k\leq e'_k < 2k-1\}\,,
$$
by shifting and relabeling the indices as $e_{n+k} := e'_k + 2n$. Note however
that $E'_{N-n}$ (with all $e'_k < 2k-1$) is {\em not} the same as $E_{N-n}$ (with
all $e_k \leq 2k-1$). 

From Eq.~\eqref{e:Fpq},
\begin{equation}
F_N(p,q)= \sum_{\{E\}}\,\prod_{k=1}^N\,[2k-e_k]_{p,q} = 
\sum_{n=0}^{N-1} F_n(p,q)C_{N-n}(p,q)\,,
\label{decomp}
\end{equation}
where $C_n(p,q)= \ds\sum_{\{E'_n\}}\ds\,\prod_{k=1}^{n}\,[2k-e'_k]_{p,q}$. Introduce 
the power series
\begin{gather*}
F(p,q,x) = \sum_{N=0}^{\infty} F_N(p,q)x^N \,, \quad F_0(p,q):= 1\,, \\
C(p,q,x) = \sum_{N=1}^{\infty} C_N(p,q)x^N \,.
\end{gather*}
Using Eq.~\eqref{decomp} in the power series, one finds that
$F(p,q,x) - 1$ equals the product $F(p,q,x)C(p,q,x)$, which implies
\begin{equation}
F(p,q,x) = \frac{1}{1-C(p,q,x)} \,.
\label{FC}
\end{equation}
Next, define the associated polynomials 
$$ F_n(p,q;l) := \sum_{\{E_n\}}\prod_{k=1}^n[2k-e_k+l]_{p,q}\,, \quad 
C_n(p,q;l) := \sum_{\{E'_n\}}\prod_{k=1}^{n}[2k-e'_k+l]_{p,q}\,, $$ 
so that $F_n(p,q;0) = F_n(p,q)$ and $C_n(p,q;0) = C_n(p,q)$.
The corresponding power series $F(p,q;l,x)$ and $C(p,q;l,x)$ are defined
similarly to $F(p,q,x)$ and $C(p,q,x)$ above, and they also satisfy
Eq.\eqref{FC}. Furthermore, for $n > 1$ the associated polynomials satisfy the relation
\begin{gather}
C_n(p,q;l) = \sum_{\{E'_n\}}\prod_{k=1}^{n}[2k-e'_k+l]_{p,q} =
[l+1]_{p,q} \sum_{\{E'_n\}}\prod_{k=2}^{n}[2k-e'_k+l]_{p,q} \\
 = [l+1]_{p,q} \sum_{\{E_{n-1}\}}\prod_{j=1}^{n-1}[2j-e_j+(l+1)]_{p,q}
= [l+1]_{p,q} F_{n-1}(p,q;l+1)\,, 
\label{Cnpql}
\end{gather}
after an appropriate index shift, $k = j+1$, and relabelings $e'_{j+1} = e_j + 1$ so that
$ j \leq e_j \leq 2j-1$ for $j=1,\ldots n-1$. As a result, the set $E'_n$ changed
to the set $E_{n-1}$. The formal power series constructed from the first and last 
expressions in Eq.\eqref{Cnpql} satisfies $C(p,q;l,x) = x[l+1]_{p,q} F(p,q;l+1,x)$. 
From the analogue of Eq.~\eqref{FC} for the associated functions $F(p,q;l,x)$ and $C(p,q;l,x)$,
one therefore obtains
$$ F(p,q;l,x) = \frac{1}{1-[l+1]_{p,q} x F(p,q;l+1,x)} \,. $$
This yields the continued fraction representation for $F(p,q,x)=F(p,q;0,x)$.
\end{proof}
We can graphically illustrate the results of Proposition~\ref{P:gen} for 
$N=2$ and 3 in terms of the corresponding chord diagrams.
For $N=2$, $F_2(p,q)=1+p+q$, which implies that there is one each
of the O-, T-, and P-type diagrams. These were displayed in Fig.~\ref{f:2sol}.
For $N=3$, there are 15 chord diagrams, which are displayed in Fig.\ref{f:N3chord}. 
They are characterized by
\[
F_3(p,q)=(1+2p+p^2+p^3)+(2+2p+2p^2)q+(1+2p)q^2+q^3\,.
\]
Note that the total number of pairwise chord configurations for each case
for $N=3$ is 3(3-1)/2 = 3. We use the ordering 
$({\bf p}_1{\bf p}_2,\, {\bf p}_2{\bf p}_3,\, {\bf p}_3{\bf p}_1)$
for the chord-pairs, and indicate the interaction type for each pair below:
\begin{enumerate}
\item[(a)] Non-resonant cases:\,\,
1 (PPP)-type, 1 (PPO)-type, 2 (POO)-type, 1 (OOO)-type. These are on 
the first row in Fig.\ref{f:N3chord}, from left to right.
\item[(b)] One-resonant cases: \,\,
2 (TPP)-type, 2 (TPO)-type, and 2 (TOO)-type. These are on the second row.
\item[(c)] Two-resonant cases:\,\, 2 (TTP)-type and 1 (TTO)-type.
These are on the the third row.
\item[(d)] Three- (i.e., fully-) resonant case: 1 (TTT)-type.
\end{enumerate}
\subsection{Generating function and $q$-orthogonal polynomials}
In the special case when $p=1$, the formula for $F(q,x):=F(1,q,x)$ in
Proposition~\ref{P:gen} reduces to similar continued fraction expression
for $F(q,x) := F(1,q,x)$, namely
\begin{equation}
F(q,x)= \sum_{N=0}^{\infty}F_N(q)x^N =\cfrac{1}{1-
                                            \cfrac{x\,[1]_{q}}{1-
                                             \cfrac{x\,[2]_{q}}{1-
                                              \cfrac{x\,[3]_{q}}{1-\cdots}}}} \,,
\qquad F_0(q):=1\,,
\label{e:Fq}
\end{equation}
with $[n]_q:= 1+q+\ldots+q^{n-1}$. From Corollary~\ref{c:specialpq}(iii), 
it follows that $F(q,x)$ is the generating
function for the polynomials $F_N(q)$ that enumerate the crossings
of the chords of $\mathcal{M}_N$, whose explicit formula is given by
Eq.~\eqref{e:TR}. Here we show that the continued fraction for $F(q,x)$ is
related to the moment generating function for the continuous $q$-Hermite
polynomials. It follows that the Touchard-Riordan polynomials 
$F_N(q)$ are simply the even moments of the weight function with respect to 
which the $q$-Hermite polynomials are orthogonal. The latter result was also
found in Ref.~\cite{EJC87}.

We first collect some facts (see e.g. \cite{deift,Moser}) from the spectral
theory of bounded, real, semi-infinite
Jacobi matrices on the Hilbert space 
$$l_2(\mathbb{C}) := \{u=(u_0,u_1,u_2,\ldots)\,:\, u_i \in \mathbb{C}\,, 
\quad \sum_{k=0}^{\infty}|u_k|^2 < \infty\}\,.$$
Define the following tri-diagonal matrices
\begin{equation*}
L:=\begin{pmatrix}
0 & 1 & 0 & \cdots   \\
a_1 &  0  & 1 & \cdots  \\
0 & a_2 & 0 & \ddots \\
\vdots  &  \ddots & \ddots & \ddots
\end{pmatrix} \quad
L_n := \begin{pmatrix}
0 & 1 & 0 & \cdots & 0 \\
a_1 &  0  & 1 & \cdots & 0 \\
 \vdots  &  \ddots & \ddots & \ddots     & \vdots \\
  0 &    \cdots   &   a_{n-2} &  0 & 1  \\
 0  &     \cdots    &      0       &    a_{n-1}   &  0
\end{pmatrix} \quad
\widehat{L}_{n-1}:= \begin{pmatrix}
0 & 1 & 0 & \cdots & 0 \\
a_2 &  0  & 1 & \cdots & 0 \\
 \vdots  &  \ddots & \ddots & \ddots     & \vdots \\
  0 & \cdots  &   a_{n-2}  &  0 &  1  \\
 0  & \cdots  &   0   &   a_{n-1}  &  0
\end{pmatrix} \,,
%\label{e:jacobi}
\end{equation*}
with $a_i > 0, \, i=1,2,\ldots$. Next, consider the linear system of equations
\[
(\lambda I -L)\phi=e_0\,, \quad {\rm for} \quad \phi = (\phi_0,\phi_1,\phi_2, \ldots)^T\,,
\]
where $I := \diag(1,1,\ldots)$ is the semi-infinite identity matrix and
$e_0=(1,0,0,\ldots) \in l_2(\mathbb{C})$.

\noindent Fact (a)\, ({\em Resolvent of Jacobi matrix}):\, The $(0,0)$-element of the 
resolvent of $L$, i.e., $\phi_0 = \langle e_0, (\lambda I - L)^{-1}e_0 \rangle$, 
can be developed in a continued fraction by rewriting the linear system of 
equations as follows:
$$
\phi_0 = \cfrac{1}{\lambda- \cfrac{\phi_1}{\phi_0}} \qquad {\rm and} \qquad
\frac{\phi_n}{\phi_{n-1}} = \cfrac{a_n}{\lambda-\cfrac{\phi_{n+1}}{\phi_n}}\,,
\quad n \geq 1 \,.$$
Thus one has
\begin{equation}
 \phi_0 = \cfrac{1}{\lambda-
                           \cfrac{a_1}{\lambda-
                                       \cfrac{a_2}{\lambda-
                                               \cfrac{a_3}{\lambda-\cdots}}}} \,. 
\label{e:phi0}
\end{equation}
For $n \geq 0$, the $n^{\rm th}$ convergent of this continued fraction 
is of the form $\ds R_n = \frac{N_n}{D_n}$, where $N_0 = 0,\, D_0 = 1$ and
$N_n(\lambda),\,\,D_n(\lambda) \,\, n \geq 1$, are polynomials in $\lambda$ of
degrees $n-1$ and $n$, respectively. These sequences of polynomials satisfy the 3-term
recurrence relation
\begin{equation}
\lambda P_n = a_n P_{n-1} + P_{n+1}\,, \qquad P_n := (N_n, \, D_n) \,,
\label{e:rec}
\end{equation}
with $P_0 = (0,1)$ and $P_1 = (1,\lambda)$ as respective initial conditions.
Furthermore, it can be shown by Cram\'er's rule that for $n \geq 2$,
$$ D_n = \det(\lambda I_n - L_n),\quad N_n = \det(\lambda I_{n-1}-\widehat{L}_{n-1}),
\quad {\rm such \,\, that} \quad \frac{N_n}{D_n} = 
\langle e_0',\, (\lambda I_n - L_n)^{-1} e_0'\rangle \,,$$
where $I_n$ is the $n\times n$ identity matrix and $e_0'=(1,0\ldots,0) \in \mathbb{R}^n$.

\noindent Fact (b)\, ({\em Spectral theorem}):\,
If $\{a_n\}, \, n \geq 1$, is a positive bounded sequence such that the
Jacobi matrix $L$, defined above, is bounded on $l_2(\mathbb{C})$, then there 
exists a unique spectral measure $\mu$ with compact support $\Sigma$ such that
\begin{equation}
\phi_0 = \lim_{n \to \infty} \frac{N_n}{D_n} = \langle e_0,\, (\lambda I - L)^{-1}e_0 \rangle
= \int_{\Sigma}\frac{d \mu(s)}{\lambda -s}\,, \quad{\rm for}\quad  \lambda \notin \Sigma\,.
\label{e:mu}
\end{equation}
Furthermore, the polynomials $\{N_n,\, D_n\},\, n\geq 1$, are orthogonal with respect
to the measure $\mu$. In particular $\{D_n\},\, n\geq 1$ form a sequence of monic
polynomials satisfying the orthogonality relations
$$ \int_{\Sigma}D_m(s)D_n(s)\,d\mu(s) = \alpha_n\delta_{mn}\,,
\quad {\rm with}\quad \alpha_n = \prod_{j=1}^na_j \,.$$

If we now set $a_n = [n]_q$ for $n \geq 1$ in the continued fraction representation
for $\phi_0$ in Eq.~\eqref{e:phi0},
and compare the resulting expression with the generating function in Eq.~\eqref{e:Fq},
we find that
$$ F(q,x=\lambda^{-2}) = \sum_{N=0}^{\infty}\frac{F_N(q)}{\lambda^{2N}}
= \lambda \phi_0(\lambda) =
\sum_{k=0}^{\infty} \frac{1}{\lambda^k}\int_{\Sigma}s^k \,d\mu(s)\,,
$$
where the last equality follows from Eq.~\eqref{e:mu}. Note also from Eq.~\eqref{e:mu}
that the moments 
$\int_{\Sigma}s^k d\mu(s) = \langle e_0,\, L^k e_0 \rangle, \,\, k=0,1,2,\ldots$,
clearly vanish for any odd $k$ because of the structure of the Jacobi matrix $L$.
Therefore, we conclude that the generating polynomials $F_N(q)$ are given by the even
moments of the measure $\mu$. Moreover, Eq.~\eqref{e:rec} with $a_n = [n]_q$ is the
well known 3-term recurrence relation for the $q$-Hermite polynomials $H_n(s,q)$ (see
e.g.,~\cite{ismail}). Indeed, it follows from the initials conditions that the
denominator polynomials above satisfy $D_n(s) = H_n(s,q)$ for $n=0,1,2,\ldots$. These
polynomials satisfy the orthogonality relations
\begin{gather*}
\int_{-a}^a\, H_m(s,q)H_n(s,q)d\mu(s) = [n]_q! \,\delta_{mn}\,, \\
d\mu(s,q) := \nu(s,q)ds\,, \qquad s \in [-a,a]\,, \qquad a:=2(1-q)^{-1/2}\,,
\end{gather*}
where $|q| < 1$, and the weight function $\nu(s,q)$ is given in terms of the
Jacobi theta function $\Theta_1(\theta, q)$ by 
\begin{gather*}
\nu(s,q) = \frac{q^{-\frac{1}{8}}}{\pi a} \Theta_1(\theta, q) =
\frac{2}{\pi a} \sum_{n=0}^{\infty} (-1)^nq^{n(n+1)/2}\sin(2n+1)\theta \,, \\
\cos \theta := \frac{s}{a} = \half (1-q)^{1/2} s\,.
\end{gather*}
Note that  $\nu(s,q)$ is an even function in $s$
(i.e., it is stable under $\theta \to \pi-\theta,$ where $ \theta \in [-\pi,\, \pi]$), 
so the odd moments vanish as observed above. The even moments for the $q$-Hermite polynomials
are given by
\begin{eqnarray}
F_N(q) & = & \int_{-a}^a x^{2N}\nu(q,x)dx 
\nonumber \\
 & = & \frac{2^{2N}}{\pi(1-q)^N} \sum_{n=0}^{\infty} (-1)^nq^{n(n+1)/2} 
\int_0^{\pi}\cos^{2N}\theta(\cos2n\theta - \cos2(n+1)\theta)\,d\theta\,,
\nonumber
\end{eqnarray}
which yields the Touchard-Riordan formula, after evaluating the last integral and
rearranging the summation indices.
\begin{remark} It is intriguing to note that the generating function 
$F(p,q,x)$ that enumerates the interaction types of the $N$-soliton solution 
of the KPII equation has its origin in the theory of $q$-orthogonal polynomials. 
We mention another relation with orthogonal polynomials, without presenting the details.
Consider the case $p=0$. It turns out that the generating function $F(0,q,x)$ for 
the non-nesting chord diagrams is related to the moment generating function for a 
certain class of $q$-orthogonal polynomials studied in Ref.~\cite{ismail83} 
(see also ~\cite{ismail}). A particularly interesting 
consequence of this relation is that the function $F(0,q,-q)$ has a 
Rogers-Ramanujan interpretation. Let $\phi_1$ and $\phi_2$ be certain 
modular forms of weight $\frac{1}{5}$ for the level-5 principal modular
group $\Gamma(5) < PSL(2,\mathbb{Z})$; namely,
\begin{equation}
\phi_1(q) =\ds{\frac{1}{\eta(q)^{3/5}}\sum_{n\in{\mathbb Z}}(-1)^nq^{(10n+1)^2/40}} \,,
\qquad \qquad
\phi_2(q)=\ds{\frac{1}{\eta(q)^{3/5}}\sum_{n\in{\mathbb Z}}(-1)^nq^{(10n+3)^2/40}}\,,
\label{e:phi12}
\end{equation}
where $\eta(q)=q^{1/24}\ds\prod_{n=1}^{\infty}(1-q^n)$ is the Dedekind $\eta$-function. 
It is well-known that the modular forms $\phi_1$ and $\phi_2$ admit infinite 
product representations, which constitute the Rogers-Ramunajan identities.
Accordingly, $F(0,q,-q)$ can be represented as a quotient:
\[
F(0,q,-q)\,= \, q^{-1/5}\frac{\phi_2}{\phi_1} \,
= \, \displaystyle{\prod_{n=0}^{\infty}\frac{(1-q^{5n+1})(1-q^{5n+4})}{(1-q^{5n+2})(1-q^{5n+3})}}\,
\, = \, \cfrac{1}{1+
                                            \cfrac{q}{1+
                                             \cfrac{q^2}{1+
                                              \cfrac{q^3}{1+\cdots}}}} \,\,.
\]
This is the Rogers-Ramanujan continued fraction~\cite{ramanujan,rogers}.
\end{remark}
%%%%%%%%%%%%%%%%%%%%%%%%%%%%%%%%%%%%%%%%%%%%%%%%%%%%%%%%%%%%%%%%%%%%%%%%%%%%%%%%
\section{Conclusion}

We have presented a classification scheme for the $N$-soliton
solutions of the KPII equation, based on the combinatorics of chord diagrams
consisting of $N$ chords connecting distinct pairs of $2N$ points. We have shown that
it is possible to associate the O-, T-, and P-type of 
pairwise interaction patterns among the $N$ asymptotic line solitons of the $N$-soliton 
configuration with the alignments, crossings and nestings among pairs of chords in the
chord diagram. As a result, the equivalence classes of $N$-soliton solutions can be 
enumerated by the same generating polynomial $F_N(p,q)$ (Eq.~\eqref{e:Fpq})
of the distribution of nestings and crossings for the set ${\mathcal M}_N$ of all 
chord diagrams of $[2N]$. It follows from Propositions 2.5 and 2.8 that each  
asymptotic line soliton of a given $N$-soliton solution is uniquely identified with an 
index pair ${\bf p}_n = [e_n, g_n]\,, \, 1 \leq e_n < g_n \leq 2N$, which also labels a 
particular chord in a chord diagram. This pairing map (Eq.~\eqref{e:pair}) plays a crucial 
rule in establishing a correspondence between an $N$-soliton equivalence class 
and a particular chord diagram of ${\mathcal M}_N$. The soliton pairing encoded in
Proposition 2.5 can be derived via a systematic asymptotic analysis (for fixed $t$) of the 
$N$-soliton $\tau$-function, which is sum of real exponentials with positive coefficients, 
by identifying those phase combinations $\Theta(m_1,\ldots,m_N)$ that are dominant in 
different regions of the $xy$-plane as $|y| \to \infty$. Since a discussion of the 
asymptotics of the $\tau$-function is beyond the main focus of the present article, 
it has been omitted here. Interested readers may find the relevant details, including 
a proof of Proposition 2.5, in ~\cite{BC} (see also ~\cite{jphysa36p10519}).

Finally, we have derived a continued fraction representation for the generating function
$F(p,q,x)$ of the polynomials $F_N(p,q)$. We have shown that the special cases of this 
generating function are, in fact, moment generating functions of certain kinds of 
$q$-orthogonal polynomials. It is interesting to speculate whether the unrestricted
$N$-soliton generating function $F(p,q,x)$ is the moment generating function for 
some new family of $(p,q)$-orthogonal polynomials.

%%%%%%%%%%%%%%%%%%%%%%%%%%%%%%%%%%%%%%%%%%%%%%%%%%%%%%%%%%%%%%%%%%%%%%%%%%%%%
\section*{Acknowledgments}

The authors would like to thank Boris Pittel of Ohio State University for his help
in proving Proposition 3.5. 

%%%%%%%%%%%%%%%%%%%%%%%%%%%%%%%%%%%%%%%%%%%%%%%%%%%%%%%%%%%%%%%%%%%%%%%%%%%%%%%
\catcode`\@ 11
\def\journal#1&#2,#3 (#4){\begingroup \let\journal=\d@mmyjournal {\frenchspacing\sl #1\/\unskip\,} {\bf\ignorespaces #2}\rm, #3 (#4)\endgroup}
\def\d@mmyjournal{\errmessage{Reference foul up: nested \journal macros}}
\def\title#1{{``#1''}}
\def\@biblabel#1{#1.}
\catcode`\@ 12

\end{document}